# Field-Free Superconducting Diode Enabled by Geometric Asymmetry and Perpendicular Magnetization

*Jiaxu Li, Zijian Zhang, Shiqi Wang, Yu He, Haochang Lyu, Qiusha Wang, Bowen Dong, Daoqian Zhu, Hisakazu Matsuki, Dapeng Zhu, Guang Yang*, Weisheng Zhao*

J. X. Li, Z. J. Zhang, S. Q. Wang, Y. He, Q. S. Wang, D. Q. Zhu, G. Yang, W. S. Zhao
School of Integrated Circuit Science and Engineering, Beihang University, Beijing 100191, China
E-mail: gy251@buaa.edu.cn

H. C. Lyu, B. W. Dong
Beijing Superstring Academy of Memory Technology, Beijing 100176, China

Hisakazu Matsuki
Toyota Riken-Kyoto University Research Center (TRiKUC), Kyoto 606-8501, Japan

D. P. Zhu, G. Yang
Integrated Circuit and Intelligent Instruments Innovation Center, Qingdao Research Institute, Beihang University, Qingdao 266100, China

The superconducting diode effect (SDE), which manifests as directional, dissipationless supercurrents, is pivotal for realizing energy-efficient superconducting logic and memory technologies. However, achieving high-efficiency SDE without external magnetic fields remains a fundamental challenge. In this study, we propose a strongly enhanced, field-free SDE in Pt/Co/Nb heterostructures, enabled by the interplay of engineered geometric asymmetry and stray fields from a perpendicularly magnetized Co layer. This configuration promotes directional vortex entry and spatially selective pinning, yielding diode efficiencies that exceed all previously reported field-free values in ferromagnet/superconductor multilayers. Temperature- and field-dependent transport measurements, supported by micromagnetic simulations, reveal that the enhanced nonreciprocity results from three cooperative mechanisms: asymmetric vortex entry, localized magnetic pinning, and Lorentz-force imbalance. These findings establish a CMOS-compatible platform for high-performance superconducting rectifiers, offering new opportunities for cryogenic spintronics and quantum electronics.

## 1. Introduction

The superconducting diode effect (SDE)[1–6], characterized by nonreciprocal critical currents, enables dissipationless supercurrent in one direction while exhibiting finite resistance in the opposite direction. This nonreciprocal behavior has significant potential for energy-efficient superconducting rectification, quantum circuits, and scalable superconducting computing systems[7–10]. However, realizing SDE requires the simultaneous breaking of time-reversal symmetry (TRS) and spatial inversion symmetry (SIS). TRS is typically broken via the application of external magnetic fields[11,12], induced proximity magnetism from ferromagnetic layers[10,13,14], or by trapping Abrikosov vortices[15,16]. SIS breaking is commonly achieved through structural asymmetry, including device-level geometry[4,7,10,17], intrinsic crystallography[1,18], or intentionally introduced localized defects[19,20].

In type-II superconductors, vortex dynamics is essential in determining the onset of dissipation. Magnetic flux penetrates the material as quantized vortices, and their motion under current induces resistive losses. Controlling vortex motion through geometric or magnetic inhomogeneities is thus essential to preserve superconductivity. Previous studies have demonstrated enhanced nonreciprocal transport by engineering asymmetric vortex pinning landscapes[4,7,10,17,21]; stray fields from adjacent ferromagnetic layers offer a means to break TRS and control vortex entry without applying external magnetic fields[10,13,20,22].

Despite these advances, a comprehensive understanding of the effect of spatially graded stray fields on vortex dynamics and their role in enhancing field-free SDE efficiency is lacking. In particular, the effect of local stray field gradients (rather than uniform field magnitude) on vortex entry, pinning, and depinning has not been systematically explored. Moreover, achieving high-efficiency, field-free diode functionality in CMOS-compatible systems remains a critical challenge. Recent demonstrations in unconventional platforms, such as moiré graphene superlattices[23–25] and multiterminal Josephson-junction networks[26,27], have achieved extremely high, sometimes near-ideal (~100%) efficiencies. However, they depend on complex heterostructures, precise twist-angle control, or multi-terminal gating, limiting scalability and CMOS compatibility.

In this study, we addressed these issues by designing ferromagnet/superconductor (FM/SC) heterostructures that combine engineered geometric asymmetry with perpendicular magnetic anisotropy (PMA). The triangular geometry introduces directional asymmetry in vortex entry barriers, and the PMA-Co layer generates strong, spatially graded out-of-plane stray fields ($B_z$), which serve as pinning and depinning modulators. Through detailed experiments and micromagnetic simulations, we identified three cooperative mechanisms underpinning the observed high-efficiency SDE: (i) preferential vortex entry on the symmetric side and (ii) enhanced vortex pinning on the asymmetric side. Two manifestations of a single asymmetric Bean–Livingston surface barrier, together with (iii) Lorentz-force asymmetry, which favors vortex depinning under reverse bias. We achieved a field-free superconducting diode efficiency of 54%, representing the highest value reported within simple, fully CMOS-compatible FM/SC multilayers fabricated by standard thin-film processes, without moiré systems or multiterminal architectures. These results demonstrate the key role of spatially graded $B_z$ fields and establish PMA-Co-based heterostructures as a robust, scalable platform for high-performance cryogenic logic and memory applications.

## 2. Results and Discussions

The proposed FM/SC multilayer stack comprised substrate//Ta(2 nm)/Pt(3 nm)/Co(0.9 nm)/Nb(40 nm) layers. This configuration was selected to ensure robust PMA (see Supplementary Note S1) while preserving continuous superconducting properties. We fabricated two types of device geometries:

- **Sample 1 (asymmetric geometry):** A superconducting strip with a triangular protrusion on one side to intentionally break structural inversion symmetry. The strip measured 100 µm in length and 5 µm in width, with the protrusion having a base of 10 µm and a height of 3.5 µm.
- **Sample 2 (symmetric geometry):** A reference Hall bar device of identical dimensions but without any geometric asymmetry.

Figure 1a shows a scanning electron micrograph of the asymmetric device alongside a schematic illustrating its operating principle. Figure 1b shows the resistance–temperature (*R*–*T*) characteristics of both symmetric and asymmetric devices. Both samples exhibit sharp superconducting transitions, with the symmetric device displaying a slightly higher critical temperature ($T_c$ = 6.5 K) than its asymmetric counterpart ($T_c$ = 6.3 K). This small difference may have resulted from local strain or edge perturbations introduced by the protrusion, slightly suppressing superconductivity at the edge without compromising overall film uniformity.

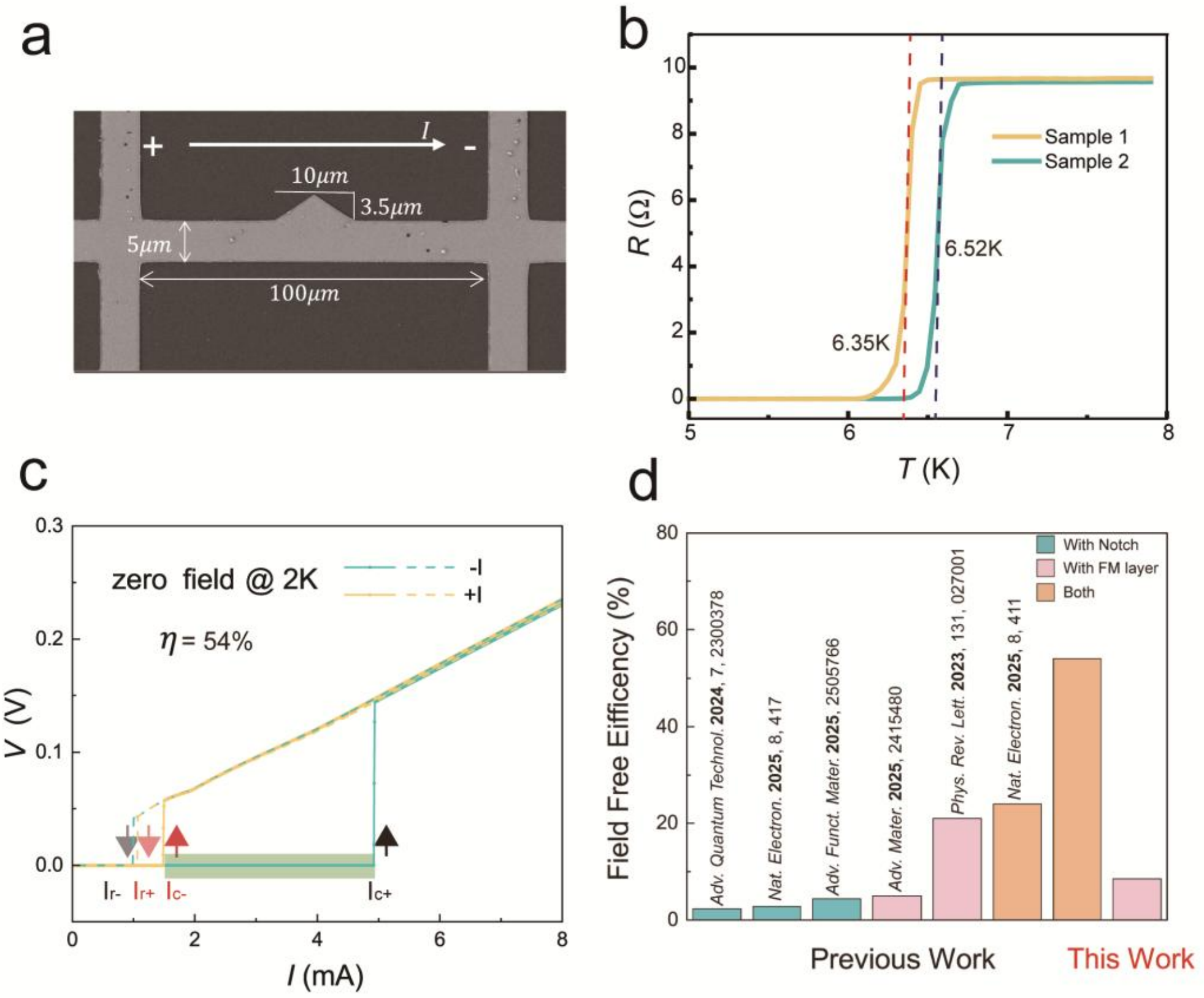

**Figure 1. Device geometry and transport behavior of symmetric and asymmetric superconducting diodes.** (a) Scanning electron micrograph of the asymmetric device (Sample 1), showing a triangular protrusion on the upper edge. The white arrow indicates the direction of current flow. (b) *R*–*T* curves of symmetric and asymmetric devices measured from 5 to 8 K under a constant current of 100 μA, showing sharp superconducting transitions and a small difference in $T_c$. (c) Current–voltage (*I*–*V*) characteristics of the asymmetric device at 2 K under zero magnetic field. Arrows indicate the direction of voltage response under forward and reverse bias. (d) Comparison of field-free superconducting diode efficiency $\eta$ of our devices and previously reported results across various material systems and device configurations. All data are measured at 2 K under zero external field after pre-magnetizing the Co layer with 200 Oe. Device width = 5 μm and length = 100 μm.

Figure 1c shows the *I*–*V* characteristics of the asymmetric device (Sample 1) measured at 2 K under a zero magnetic field. The positive ($I_c^+$) and negative critical currents ($I_c^-$) reaches 4.92 and 1.48 mA, respectively, yielding a diode efficiency of $\eta = \frac{I_c^+ - [I_c^-]}{I_c^+ + [I_c^-]} = 54\%$. This value significantly exceeds previously reported results for comparable FM/SC systems[13,14,20]. In contrast to previous notch- or edge-asymmetry devices, which demonstrate vortex-driven nonreciprocity[28–31] but often operate under finite magnetic fields or in platforms less compatible with large-scale CMOS processing, our device incorporates field-free, CMOS-friendly FM/SC multilayers fabricated entirely by standard thin-film techniques. Figure 1d benchmarks the superconducting diode efficiencies of the asymmetric device (Sample 1) and the symmetric control device (Sample 2), each 5-μm wide and 100-μm long, measured at 2 K under a zero external field after a 200 Oe perpendicular pre-magnetization of the Co layer, against representative implementations in conventional FM/SC systems, highlighting performance within this materials class[4,7,10,13,17,20]. Notably, Sample 1 achieves the highest field-free diode efficiency reported for scalable FM/SC multilayer devices fabricated using CMOS-friendly processes; the symmetric reference device exhibits substantially lower efficiency under identical zero-field conditions, emphasizing the critical function of inversion-symmetry breaking. The enhanced nonreciprocity originates from asymmetric vortex dynamics induced by the triangular protrusion. This nonreciprocity modifies local surface barriers and promotes current crowding. In addition, the perpendicularly magnetized Co layer generates intrinsic stray fields that distort the Meissner state and facilitate vortex penetration. This combined geometric and magnetic asymmetry amplifies directional vortex motion, establishing the foundation for the highly efficient field-free SDE observed in the proposed devices.

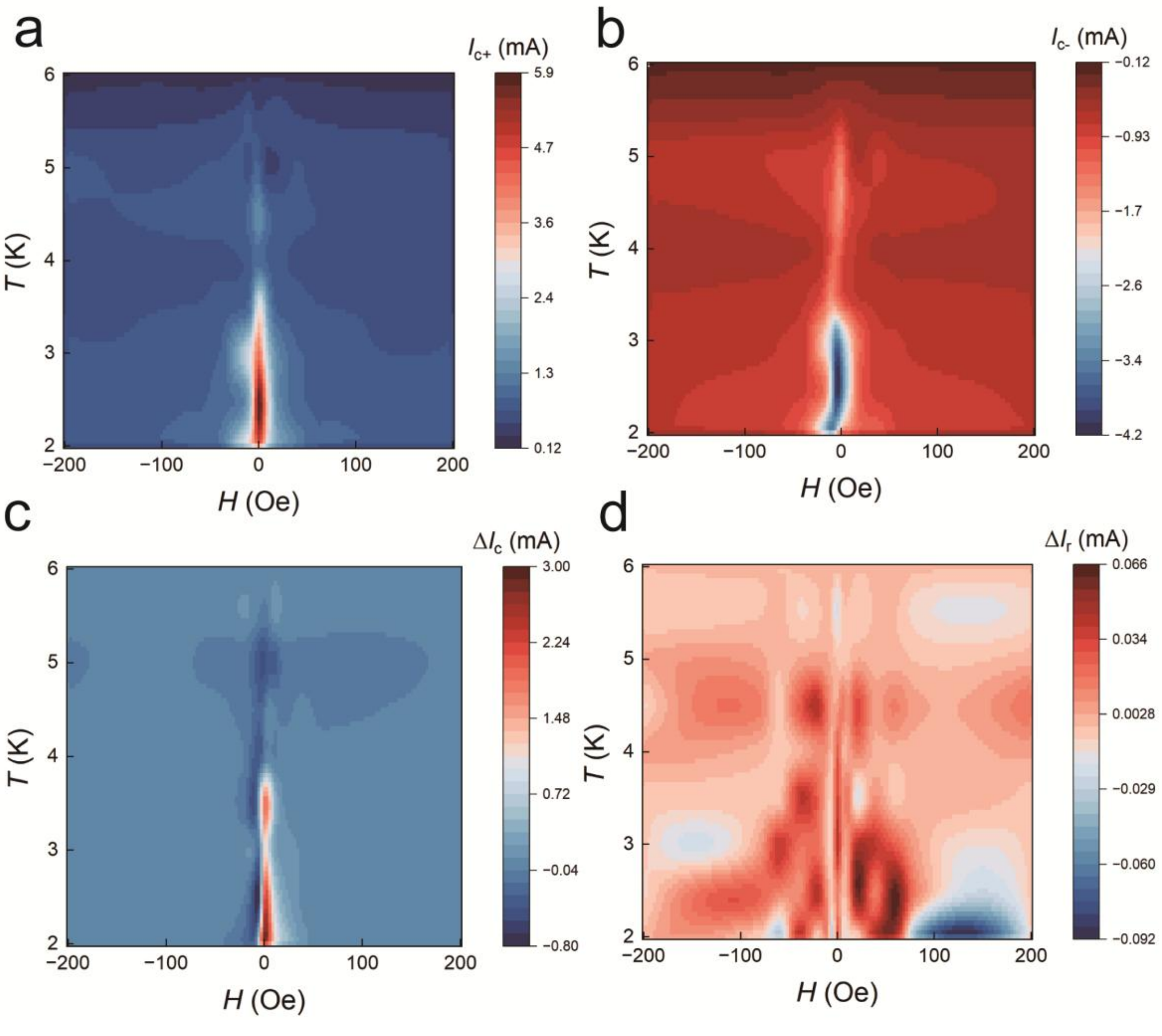

**Figure 2. Temperature- and field-dependent evolution of critical currents and nonreciprocal components in the asymmetric superconducting diode (Sample 1).** (a) Colormap of the positive critical current $I_c^+$ as a function of temperature and out-of-plane magnetic field. (b) Colormap of the negative critical current $I_c^-$ under identical conditions. (c) Corresponding nonreciprocal component $\Delta I_\mathrm{c} = I_\mathrm{c}^+ - | I_\mathrm{c}^- |$, showing a monotonic decrease with increasing temperature, and finally vanishing near $T_\mathrm{c}$. (d) Colormap of the rectified retrapping current amplitude $\Delta I_\mathrm{r}$. All measurements are performed at temperatures between 2 and 6 K under perpendicular magnetic fields up to ±100 Oe.

Vortex behavior in type-II superconductors is highly sensitive to temperature. Thermal fluctuations affect vortex pinning strength, mobility, and annihilation dynamics, all of which influence the stability of nonreciprocal transport.[32–34] To investigate this, we systematically measured the nonreciprocal current component, defined as $\Delta I_c = I_c^+ - |I_c^-|$, over a temperature range of 2 to 6 K, and similarly defined $\Delta I_\mathrm{r}$ for the retrapping currents during the resistive-to-superconducting transition. Figures 2a–d present colormaps of $I_\mathrm{c}^+$, $I_\mathrm{c}^-$, $\Delta I_\mathrm{c}$, and $\Delta I_\mathrm{r}$, respectively, as functions of temperature and out-of-plane magnetic field. Field-dependent *I–V* characteristics of Sample 1 at 2 K are provided in Supplementary Note S2. Notably, the nonreciprocal critical current is maximized at a zero magnetic field, even after reversing the magnetization direction of the Co layer (Supplementary Note S3), indicating that stray field-induced vortex asymmetry is most prominent in the absence of external fields. Introducing an external magnetic field disrupts this balance, gradually restoring symmetry in vortex dynamics and thereby reducing the diode efficiency. As shown in the $\Delta I_\mathrm{c}$ colormap, the nonreciprocal signal decreases monotonically with increasing temperature and vanishes near $T_\mathrm{c}$, consistent with the suppression of vortex dynamics at elevated temperatures and underscoring their key role in SDE efficiency. By contrast, $\Delta I_\mathrm{r}$ shows no discernible polarity dependence, confirming that the retrapping transition is thermally driven and thus unaffected by the directional vortex dynamics responsible for $\Delta I_\mathrm{c}$ asymmetry.

At low temperatures, the superconducting state is robust, vortex density remains low, and vortex–vortex interactions are weak. In this regime, local factors, such as stray fields, structural inhomogeneity, or geometric asymmetry, dominate vortex motion, resulting in strong directionality and pronounced nonreciprocity. As temperature increases, the superconducting gap narrows and vortex density rises, leading to enhanced vortex–vortex interactions, which destabilize the asymmetric configuration[35]. Simultaneously, the critical current decreases, and thermal activation becomes more prominent. Thermally-assisted vortex depinning and hopping reduce the effectiveness of pinning centers and increase vortex mobility, further diminishing the diode effect. Previous studies have shown that vortex matter undergoes a crossover from a quasi-solid phase (e.g., vortex glass or Bragg glass) to a more fluid-like state at higher temperatures[33,36–39]. In this high-temperature regime, thermal fluctuations weaken vortex pinning and erode the geometric control over vortex trajectories, ultimately suppressing the efficiency of nonreciprocal transport.

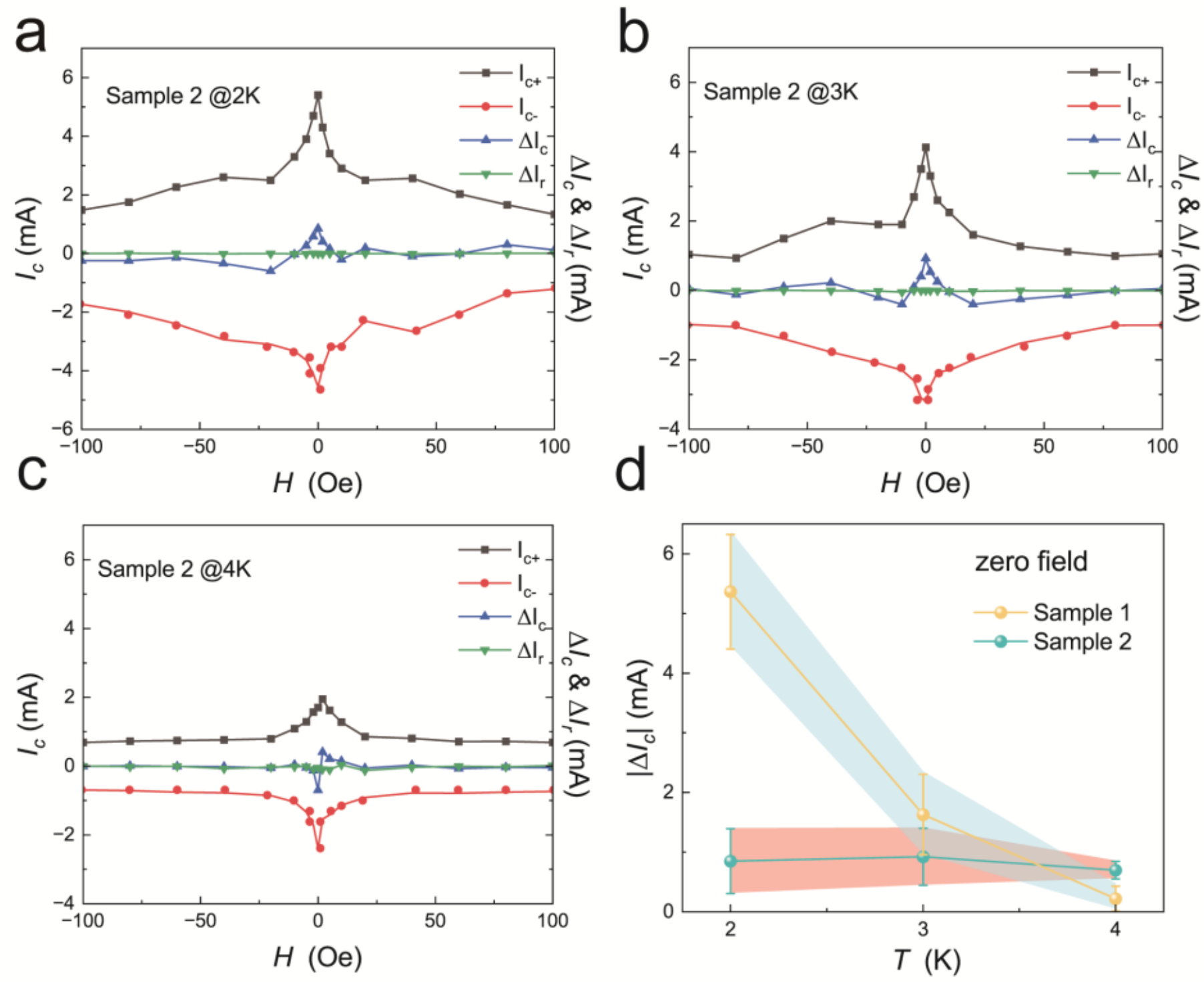

**Figure 3. Comparison of critical current behavior in symmetric and asymmetric superconducting devices.** (a)–(c) Magnetic field dependence of the critical currents $I_c^+$ and $I_c^-$ for the symmetric device (Sample 2) at 2, 3, and 4 K, respectively, showing minimal nonreciprocity across all temperatures and field conditions. (d) Comparison of the nonreciprocal current component $\Delta I_c$ under zero magnetic field for the asymmetric (Sample 1) and symmetric (Sample 2) devices at various temperatures. Colored bands represent error margins based on repeated measurements, reflecting experimental uncertainty.

We compared the *I*–*V* characteristics of the symmetric device (Sample 2) at 2, 3, and 4 K, as shown in Figures 3a–c. Although a weak diode effect is observable at all temperatures, its magnitude is significantly lower than that of the asymmetric structure. Figure 3d directly compares $\Delta I_c$ between the two devices. The asymmetric structure exhibits a considerably larger $\Delta I_c$, particularly at low temperatures. At 2 K, $\Delta I_c$ reaches 5.4 mA in the asymmetric sample, over six times higher than the 0.85 mA measured in the symmetric counterpart. This substantial enhancement arises from the directional modulation of vortex dynamics enabled by the engineered geometry. Specifically, the triangular protrusion in the asymmetric device modifies the local vortex entry barrier, promoting stronger pinning on one side while enabling easier vortex motion on the other. This anisotropic behavior results in pronounced nonreciprocal transport. By contrast, the symmetric device lacks features that break inversion symmetry, leading to a uniform vortex landscape and significantly weaker SDE. These results highlight geometric asymmetry as a key parameter for controlling vortex behavior, particularly in the low-temperature regime where thermal fluctuations are minimal. However, geometric asymmetry alone is not sufficient to account for the high efficiency observed in Sample 1. As shown by these measurements and previous results on geometry-only devices without magnetic layers[4,17], and likewise by PMA-only controls in this study, neither factor alone approaches the performance of the proposed asymmetric PMA device. The markedly enhanced SDE arises from the joint action of geometric asymmetry and perpendicular magnetization, together amplifying directional vortex motion.

In type-II superconductors, vortex dynamics are governed by the interplay between Lorentz forces and pinning forces. In the absence of an external magnetic field, Lorentz forces arise from the interaction between current and stray fields generated by the adjacent ferromagnetic layer. When the ferromagnet exhibits uniform perpendicular magnetization, it produces out-of-plane stray fields that penetrate the superconducting layer. In asymmetric structures, this field distribution becomes highly inhomogeneous, inducing asymmetric Meissner screening currents along the device edges[40]. Recent studies have identified such asymmetric screening currents as essential for achieving the SDE, particularly in systems with broken inversion symmetry[20].

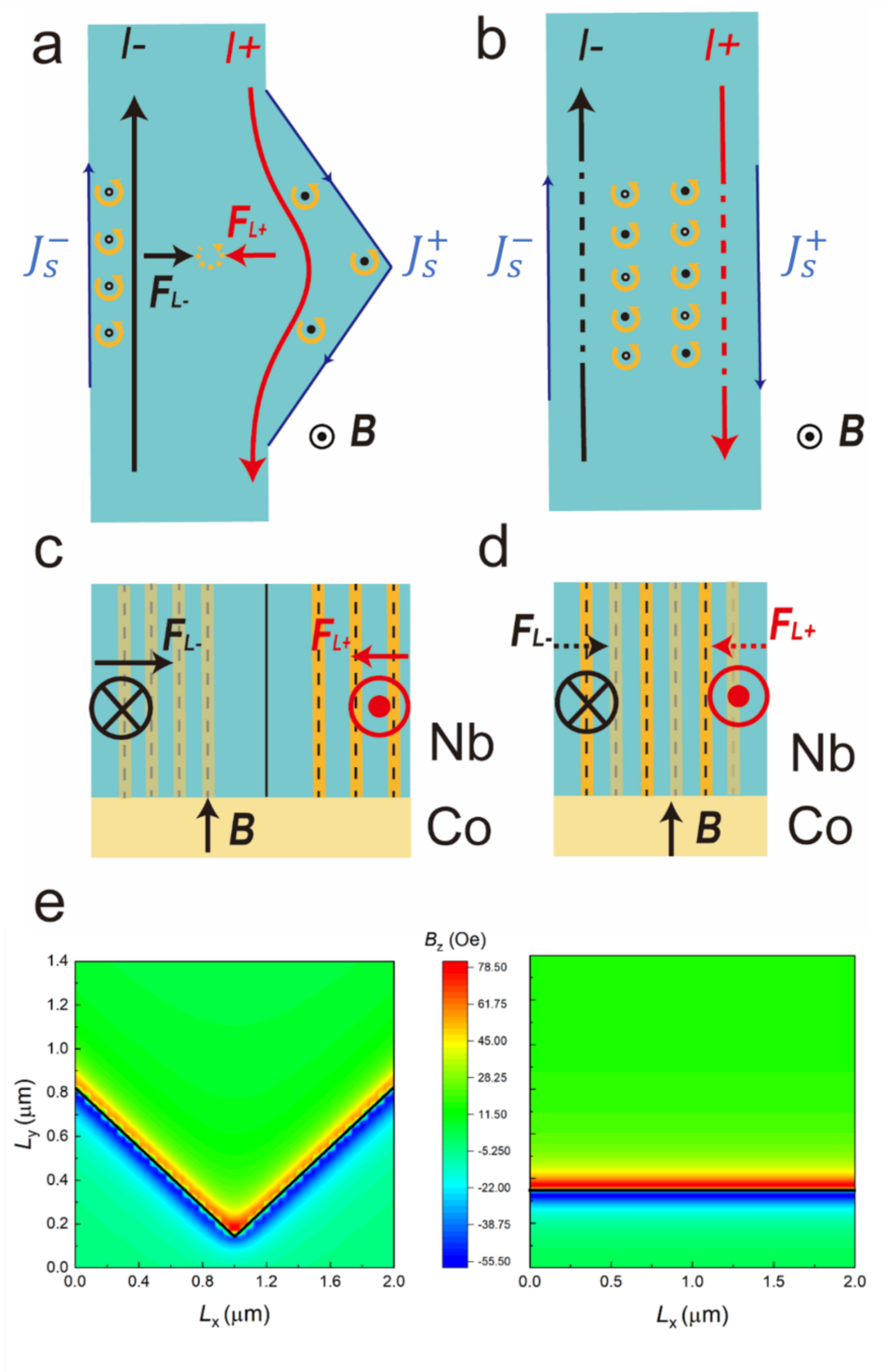

**Figure 4. Vortex dynamics and stray field distribution in asymmetric and symmetric superconducting devices.** (a) and (c) correspond to the asymmetric device. (a) Schematic top view illustrating vortex trajectories under forward and reverse current bias. Vortices are represented as ring-like structures with central dots, and blue arrows indicate Meissner screening currents. The triangular protrusion induces directional asymmetry in vortex entry. (c) Cross-sectional views under positive and negative bias, respectively. Light and dark yellow stripes denote the vortex pinning potential, and the black dashed line marks the location of the pinning center. (b) and (d) correspond to the symmetric device. (b) Schematic top view of the structure without geometric asymmetry. (d) Cross-sectional views under positive and negative bias, illustrating symmetric vortex behavior due to the uniform geometry. (e) Contour map of the local stray field distribution near the asymmetric and symmetric edge regions, highlighting the spatial variation introduced by geometric asymmetry.

Figure 4 illustrates vortex dynamics and stray field distributions in asymmetric and symmetric superconducting devices. Figure 4a shows how the triangular protrusion increases the local vortex entry barrier on the asymmetric side, thereby modulating the vortex potential landscape. This geometric asymmetry induces spatially nonuniform stray fields near the device edges, denoted as $B_{stray}^{A/S}$ for the asymmetric and symmetric sides, respectively. As shown in the MuMax3-simulated contour maps in Figure 4e, the average stray field magnitude on the asymmetric side ($B_{stray}^{A}$) is lower than that on the symmetric side ($B_{stray}^{S}$). In the absence of a bias current, the local Meissner screening currents at the asymmetric and symmetric edges can be expressed as $J_s^{+} = \alpha B_{stray}^{A}$ and $J_s^{-} = \alpha B_{stray}^{S}$, respectively, where $\alpha$ is a proportionality constant[20]. The weaker screening current on the asymmetric side leads to a higher surface barrier for vortex entry, thus suppressing vortex penetration from that edge[12,41]. Consequently, vortices preferentially enter from the symmetric side.

Micromagnetic simulations further reveal a steeper spatial gradient of the stray field near the asymmetric edge. This gradient yields localized magnetic potential wells, which serve as strong pinning centers for the few vortices that penetrate this region[42–44] (Figure 4c). Notably, the phenomena, previously described as preferential vortex entry and enhanced pinning, reflect a single mechanism: an uneven Bean–Livingston surface barrier arising from the asymmetric edge geometry and its associated pinning landscape. This combined mechanism naturally leads to a spatially asymmetric distribution of vortex dynamics and dissipation, underlying the observed nonreciprocal response. We consider the dynamic behavior under an applied current $J_{ext}$ to examine another cooperative mechanism governing the nonreciprocal response. Under current bias, the total current density in the superconducting layer is given by $J_{tot} = J_{ext} + J_s^{\pm}$. The Lorentz force acting on edge-localized vortices is given by $F_L = J_{tot} \times B_z$, where $B_z$ denotes the out-of-plane stray field produced by the Co layer. Under forward bias ($I^{+}$), this Lorentz force acts predominantly on the asymmetric edge and is given by $F_{L+} = \left(J_{ext} + \alpha B_{stray}^{A}\right) \times \phi_0$, where $\phi_0$ is the magnetic flux quantum[17,20]. Under reverse bias ($I^{-}$), the Lorentz force acts mainly on the symmetric edge, yielding $F_{L-} = \left(J_{ext} + \alpha B_{stray}^{S}\right) \times \phi_0$. As $B_{stray}^{A} < B_{stray}^{S}$, it follows that $F_{L+} < F_{L-}$. This asymmetry in Lorentz force implies that vortices on the asymmetric edge experience a weaker driving force under forward bias, rendering them less likely to be depinned from strong pinning centers. By contrast, under reverse bias, the stronger Lorentz force is more effective at addressing the weaker pinning at the symmetric edge, resulting in earlier dissipation. Consequently, a higher critical current $I_c^{+}$ is required to initiate vortex motion under forward bias, whereas a lower critical current $I_c^-$ suffices under reverse bias. This Lorentz-force asymmetry constitutes another cooperative mechanism, which, together with asymmetric entry barriers and spatially selective pinning, accounts for the high-efficiency SDE observed in the proposed devices. By contrast, Figures 4b and 4d show that the symmetric reference device lacks geometric asymmetry, resulting in balanced vortex entry, uniform pinning, and negligible directional variation in Lorentz force, which is consistent with its almost reciprocal transport behavior. To further illustrate this mechanism, we developed a field-informed quasi-static model (Supplementary Note S4) that visualizes the asymmetric Bean–Livingston barrier and reproduces the observed direction-dependent vortex occupancy and rectification. A direct real-space imaging of vortex motion under these conditions, for example, via Lorentz microscopy or magneto-optical imaging, would provide the most explicit experimental verification and is planned for future work[22,45,46].

To further elucidate the microscopic origin of the aforementioned spatially graded stray fields, we performed micromagnetic simulations comparing devices with PMA and in-plane magnetic anisotropy (IMA), as shown in Figure 5. These simulations reveal the governing influence of the interplay between device geometry and magnetic anisotropy on the formation of stray-field gradients, and highlight the essential role of PMA in enabling nonreciprocal vortex dynamics and the superconducting diode effect.

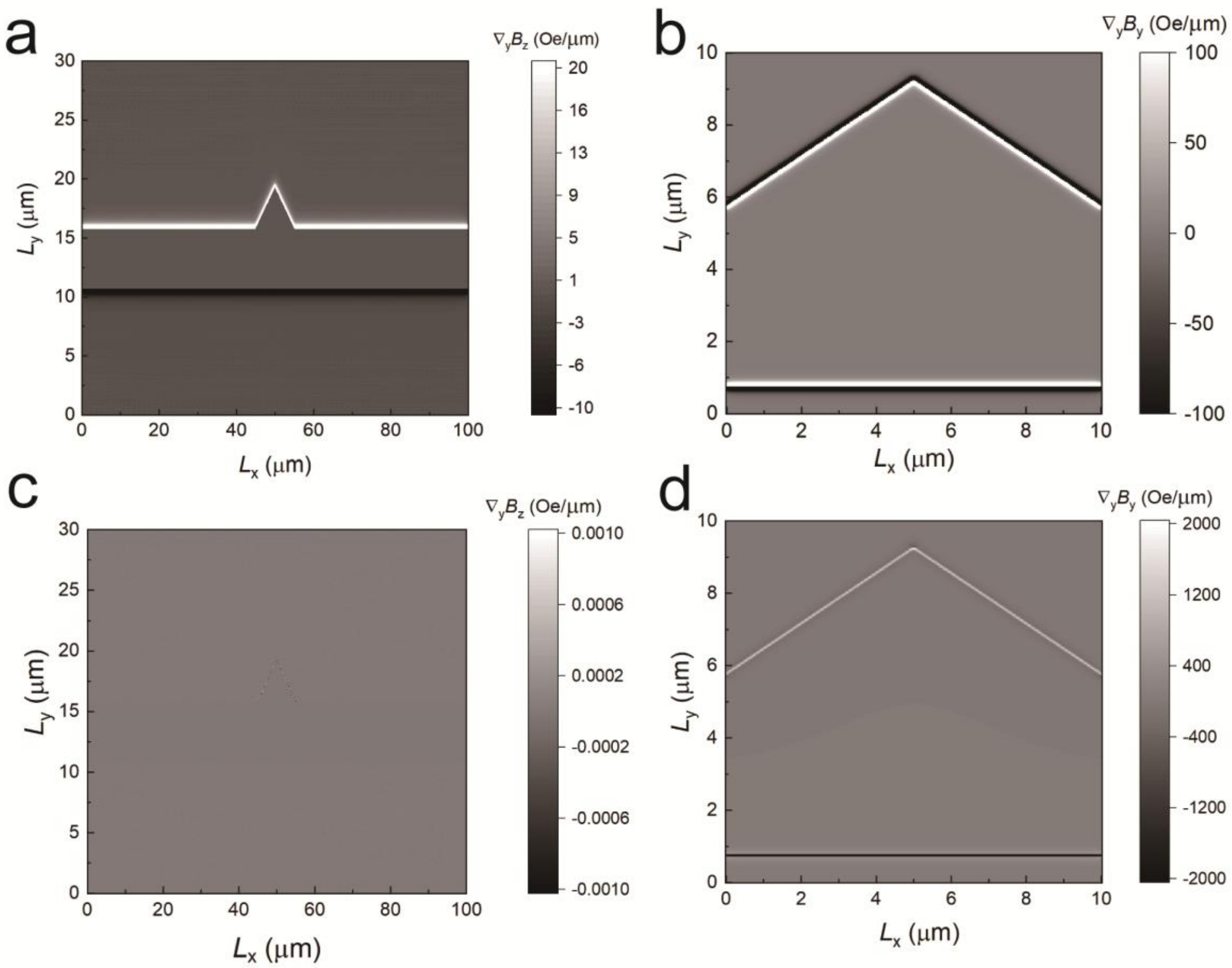

**Figure 5. Micromagnetic simulations of stray field gradients in asymmetric devices with PMA and IMA.** (a) Simulated spatial gradient of the out-of-plane stray field component $B_z$ along the y-direction at the upper and lower edges of the PMA-Co device, showing enhanced variation near the triangular protrusion. (b) Simulated gradient of the in-plane stray field component $B_y$ at a height of 40 nm above the Co layer, corresponding to the Nb surface. (c, d) Corresponding $B_z$ and $B_y$ gradient profiles for an identical geometry with IMA-Co, showing negligible $B_z$ gradients and similar $B_y$ behavior to the PMA case. These results highlight the essential role of spatially graded out-of-plane fields in enabling the SDE.

Figures 5a and 5b illustrate the simulated stray-field gradients in the PMA configuration. The out-of-plane magnetized Co layer generates substantial $B_z$ components with steep spatial gradients, particularly near the geometric transition between the rectangular and triangular regions. Figure 5a shows the gradient of $B_z$ along the y-direction at the device edges (simulated over a 100 μm × 30 μm × 1 nm region), whereas Figure 5b shows the corresponding $B_y$ gradient at 40 nm above the Co layer (i.e., the Nb surface) within a 10 μm × 10 μm × 41 nm domain. Although minor modulation of $B_y$ is observed, its spatial gradient is relatively weak and symmetric. These results confirm that in the asymmetric configuration, PMA enhances the magnitude of $B_z$ and introduces steep and localized $B_z$ gradient, both of which are critical for directional vortex control. We also note that as linewidths and spacings enter the sub-micron regime, stray-field reach can approach device separation and must be treated as a layout constraint. Practical mitigations include: (i) enforcing a minimum keep-out distance of ~4-5 μm (or as determined by stack-specific simulations); (ii) employing far-field-cancelling magnetic stacks, such as synthetic antiferromagnets with perpendicular magnetic anisotropy (SAF-PMA) or flux-closure geometries to preserve near-edge gradients while suppressing long-range fields; and (iii) incorporating superconducting shields or ground planes between elements where necessary. Overall, our simulations (Supplementary Note S5) predict negligible cross-talk at current spacings, establishing clear, CMOS-compatible design rules that ensure scalability as density increases.

To isolate the role of magnetic anisotropy, we simulated an identical geometry with in-plane magnetization by increasing the Co thickness to 2 nm and setting the uniaxial anisotropy constant $K_u = 0$. A 0.1-T external field was applied along the y-direction to uniformly align the magnetization. Compared with the PMA case, the IMA results (Figures 5c and 5d) exhibit negligible $B_z$ gradients. Although minor inhomogeneities in $B_y$ arise from magnetization nonuniformity, the lack of significant out-of-plane stray fields renders the IMA configuration ineffective for driving nonreciprocal vortex transport. These findings demonstrate that the presence of a spatially nonuniform out-of-plane stray field, not merely its magnitude, is crucial for inducing nonreciprocal vortex dynamics. Compared with the IMA configuration, PMA-Co heterostructures produce stronger $B_z$ fields and steeper spatial gradients at the device edges, which are essential conditions for realizing efficient, field-free superconducting diode operation.

Notably, intrinsic SDEs, such as those based on non-centrosymmetric superconductors or engineered Rashba systems[1,47–49], can provide dissipationless rectification at low currents but typically require complex heterostructures and offer limited post-fabrication tunability, which constrains large-scale integration. The proposed PMA-Co approach differs from previous reports on EuS-based ferromagnetic insulators, in which sizable $B_z$ fields originate from intrinsic $4f$ orbital moments, even under in-plane magnetization[50,51]. Although those systems can support SDE, their reliance on material-specific magnetic properties limits scalability and integration. By contrast, the proposed PMA-based design achieves the highest field-free superconducting diode efficiency reported for CMOS-compatible FM/SC multilayers while utilizing a fabrication-friendly, fully scalable material platform. These combined advantages establish PMA-Co heterostructures as a robust foundation for integrating high-performance superconducting diodes into next-generation cryogenic spintronic, logic, and memory architectures[52].

## 3. Conclusion

In this study, we demonstrated that combining geometric asymmetry with stray fields from a perpendicularly magnetized ferromagnetic layer enables efficient modulation of vortex dynamics, leading to a strongly enhanced SDE under zero external magnetic field. By engineering asymmetric FM/SC heterostructures, we established a microscopic link between spatial stray-field gradients, directional vortex pinning, and nonreciprocal superconducting transport. These results provide direct insight into the mechanism of field-free SDE and indicate that further enhancements may be achieved through magnetization control or advanced structural design. The robust, geometry-driven SDE presented here offers a scalable, CMOS-compatible approach to superconducting device engineering, with strong potential for low-power, non-volatile logic and memory applications in cryogenic and quantum technologies.

## 4. Experimental Section

*Sample Preparation*: All samples were fabricated on thermally oxidized silicon substrates using DC magnetron sputtering in an ultrahigh vacuum system with a base pressure below $10^{-8}$ Torr at room temperature. Device patterning was performed using electron-beam lithography with a positive photoresist, followed by electron-beam evaporation of a 120-nm-thick Al layer, which served as a hard mask. After standard lift-off in acetone, the Al mask was used for pattern transfer via ion beam etching. Residual Al was subsequently removed using tetramethylammonium hydroxide.

*Electrical and Magnetic Measurements*: Electrical and magnetic measurements were performed in a Physical Property Measurement System (PPMS, Quantum Design) using a standard four-probe configuration. DC was sourced using a Keithley 6221 current source, and voltage was measured with a Keithley 2182A nanovoltmeter. A magnetic field was applied perpendicular to the device plane unless otherwise specified. Before zero-field measurements, the Co layer was magnetized by applying a perpendicular magnetic field exceeding its coercive field (~200 Oe) to establish a remanent magnetization state. This initialization step was performed once after sample mounting and remained stable over repeated thermal cycles and throughout all low-temperature measurements. All measurements were performed in a shielded environment to minimize external noise and thermal fluctuations.

**Supporting Information**

- Supplementary Note S1. Magnetic characterization of the Co layer.
- Supplementary Note S2. Field-dependent *I*–*V* characteristics at 2 K.
- Supplementary Note S3. Magnetization-dependent $\Delta I_c$ behavior.
- Supplementary Note S4. Field informed quasi-static model of the asymmetric Bean–Livingston barrier.
- Supplementary Note S5. Stray-field crosstalk Analysis.

**Acknowledgements**

This work was supported by the National Key Research and Development Program of China (Grant No. 2022YFA1402604), the Natural Science Foundation of China (52201200, 52401225), the Taishan Scholars Program, and Research Funds for the Central Universities. The authors would like to acknowledge the Center for Micro-Nano Innovation of Beihang University for the assistance in the fabrication of the device reported in this work.

Received: ((will be filled in by the editorial staff))
Revised: ((will be filled in by the editorial staff))
Published online: ((will be filled in by the editorial staff))

**Supplementary Note S1. Magnetic characterization of the Co layer.**

To verify the magnetic properties of the Co layer, vibrating sample magnetometry (VSM) measurements were performed at room temperature, as shown in Figure S1. The hysteresis loop reveals a clear magnetic response, indicating strong perpendicular magnetic anisotropy, consistent with previous reports on ultrathin Co films[1,2].

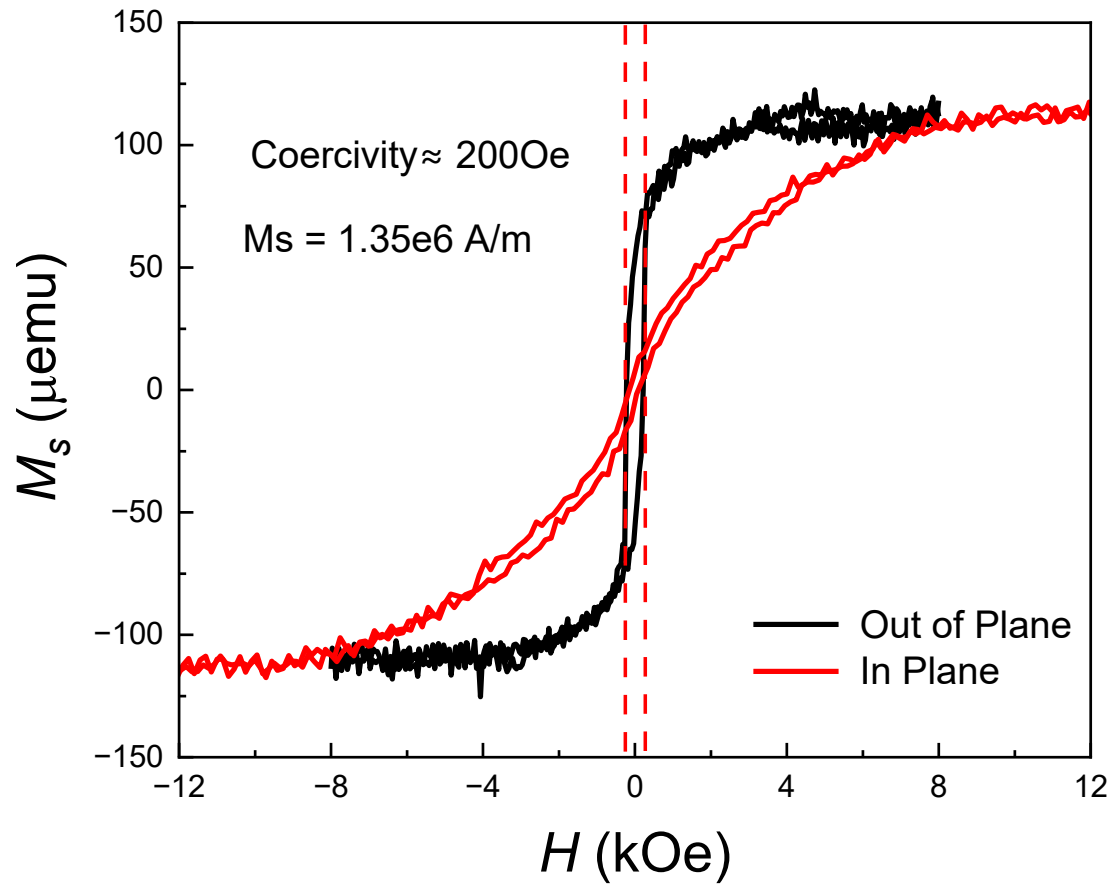

Figure S1. Magnetic hysteresis loop of the Co layer measured at room temperature.

**Supplementary Note S2. Field-dependent *I-V* Characteristics at 2 K.**

Figure S2 presents the *I-V* characteristics of Sample 1 at 2K under magnetic fields ranging from -200 Oe to 200 Oe. Solid lines indicate current sweeps from negative to positive, while dashed lines represent sweeps in the reverse direction. The critical current $I_c$ reaches its maximum at zero magnetic field. Importantly, the retrapping current $I_r$ remains nearly constant across all curves, suggesting that Joule heating has a negligible effect in this study.

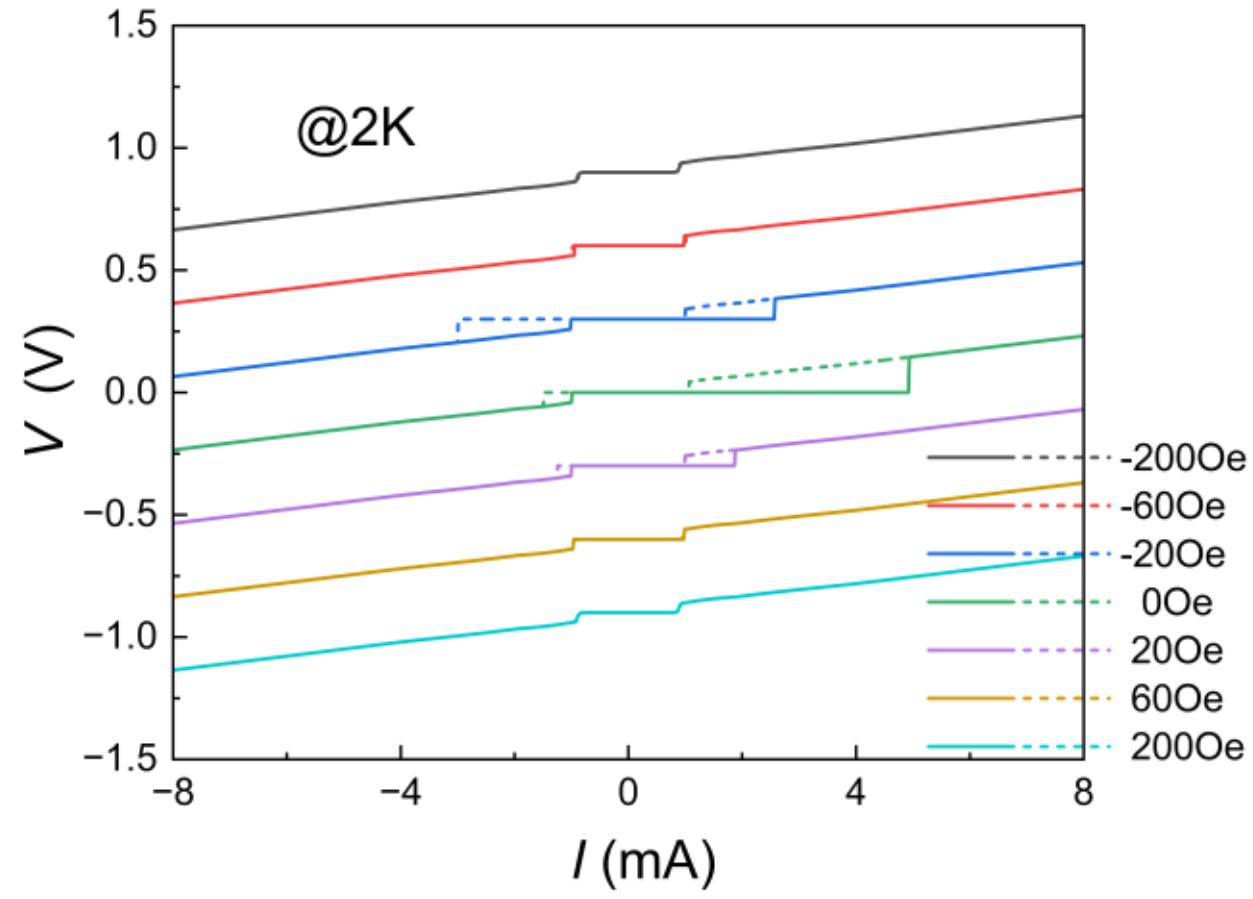

Figure S2. *I*-*V* characteristics of Sample 1 at 2 K under magnetic fields from –200 Oe to 200 Oe. Solid lines denote current sweeps from negative to positive, while dashed lines indicate sweeps in the reverse direction.

## Supplementary Note S3. Magnetization-dependent $\Delta I_c$ behavior.

Figure S3 shows the dependence of $\Delta I_c$ on the applied magnetic field after reversing the magnetization direction of the Co layer. Prior to measurement, an oscillating demagnetization process was performed under a 1 T field, followed by a +1 T or -1 T field to ensure full saturation of the ferromagnetic layer in the desired direction. As shown, $\Delta I_c$ exhibits the expected odd-symmetric reversal upon magnetization switching[3,4], highlighting the crucial role of the stray magnetic field in the SDE observed in this system.

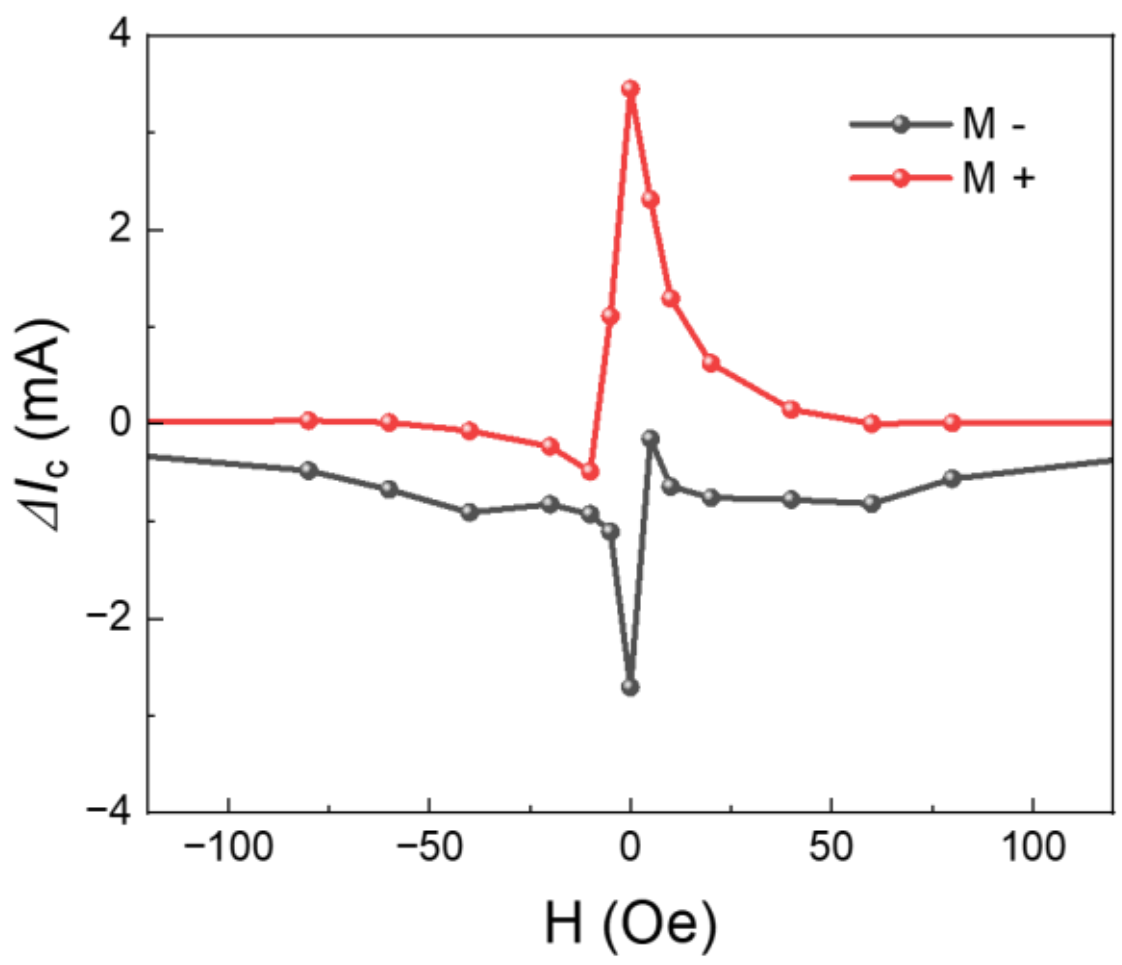

Figure S3. $\Delta I_c$-H curves at 2K after magnetic saturation following oscillating demagnetization. The red curve corresponds to the Co layer magnetized along the +*z* direction, while the black curve represents magnetization along the -*z* direction.

## Supplementary Note S4. Field informed quasi-static model of the asymmetric Bean–Livingston barrier.

To complement the experimental results and provide microscopic insight into the origin of the superconducting diode effect, we developed a field-informed quasi-static model based on the measured out-of-plane stray-field distribution $B_z(x, y)$. This model enables a direct comparison between asymmetric and symmetric device edges under forward and reverse current bias.

In this framework, the Bean–Livingston surface barrier is reconstructed from $B_z(x, y)$, and a gradient-based magnetic pinning term is added to account for local field inhomogeneity. From the resulting potential landscape, we compute zero-bias vortex occupancy maps and, under applied current bias, visualize the corresponding biased maps with representative vortex positions and trajectories. Although this quasi-static approach does not include explicit time evolution, it effectively captures the key signatures of the mechanism—preferential vortex entry, selective pinning, and current-direction rectification—within a physically consistent, mechanism-driven framework.

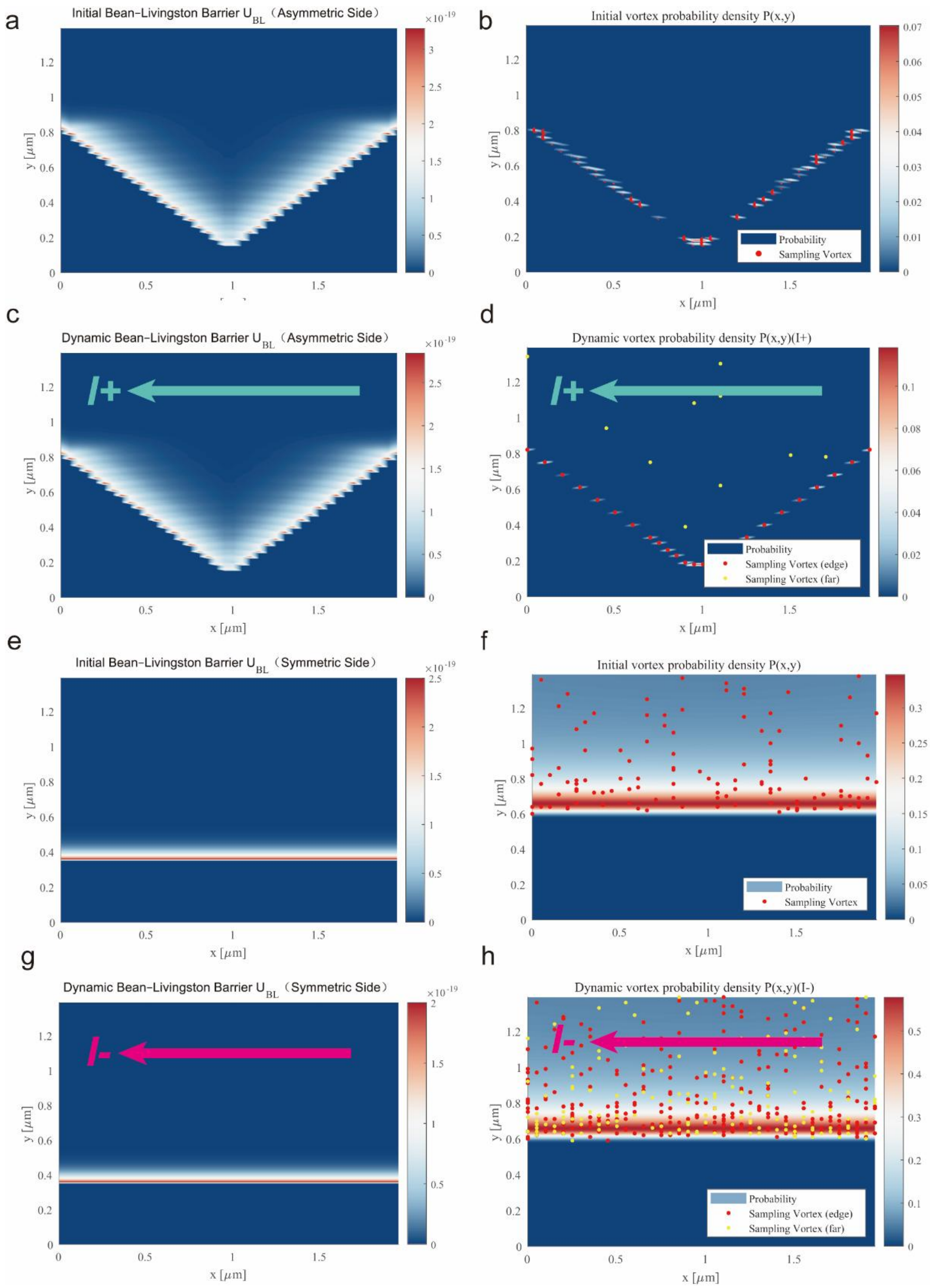

Figure S4. Field-informed minimal model of vortex rectification (quasi-static). (a-d) Asymmetric edge: (a) zero-bias $U_{BL}(x,y)$ reconstructed from $B_z(x,y)$; (b) zero-bias occupancy $P(x,y)$ with sampled entry/occupation sites (red); (c) forward-bias ($I^+$) barrier reduction near the asymmetric edge; (d) biased occupancy with current-driven vortex positions/trajectories (yellow dots). (e-h) Symmetric edge: (e) zero-bias $U_{BL}(x,y)$; (f) zero-bias $P(x,y)$ with sampled sites (red); (g) reverse-bias ($I^-$) barrier reduction near the symmetric edge; (h) biased occupancy with current-driven vortices (yellow). Arrows indicate

current direction.

Starting from $B_z(x,y)$, we reconstruct the Bean-Livingston surface barrier and add a gradient-based pinning term to obtain a field-informed potential. At each grid point,

$$U_{BL}(x,y) = \alpha \frac{B_{ref}}{|B_z(x,y)|_{edge} + \varepsilon} e^{-d(x,y)/\lambda_{eff}}$$

where $B_{ref}$ is a robust edge-field reference, $d(x,y)$ is the distance to the nearest edge, $\lambda_{eff}$ is the effective penetration depth, and $\varepsilon > 0$ avoids singularities when $|B_z|_{edge} \to 0$. Regions with weaker edge stray fields (the asymmetric side) therefore exhibit a higher surface barrier, consistent with screening currents. The gradient-derived pinning potential is:

$$U_{pin}(x,y) = -\gamma|\nabla B_z(x,y)| \cdot \left[1 + k_{pin} \cdot e^{-d(x,y)/l_{pin}}\right]$$

weighted by edge proximity to emphasize steep gradients and geometric vertices. Here $\gamma$ sets the pinning depth (energy scale), $k_{pin}$ is a dimensionless edge-enhancement factor, and $l_{pin}$ is the edge-proximity length over which the enhancement decays to $1/e$ into the interior. Thus, the zero-bias potential and occupancy are:

$$U(x,y) = U_{BL}(x,y) + U_{pin}(x,y)$$

$$P(x,y) \propto \exp\left[-U(x,y)/k_B T\right]$$

and random samples from $P(x,y)$ visualize likely entry/occupation sites.

To emulate current bias in this quasi-static framework, we write:

$$J_{tot} = J_{ext} + \alpha_J B_{stray}^{A/S}$$

and include an edge-localized barrier-reduction term acting within a narrow layer next to the edge:

$$U_L(x,y) = -\Phi_0 J_{tot} l_L e^{-d(x,y)/l_L}$$

where $\Phi_0$ is flux quantum and $l_L$ is the effective length of the Lorentz-force-equivalent deceleration action. The biased potential is then:

$$U(x,y) = U_{BL}(x,y) + U_{pin}(x,y) + U_L(x,y)$$

Forward bias ($I^+$) mainly lowers the barrier at the asymmetric edge, while reverse bias ($I^-$) does so at the symmetric edge. Because $B^A_{stray} < B^S_{stray}$, the effective drive is weaker on the asymmetric side under $I^+$ and stronger on the symmetric side under $I^-$, reproducing nonreciprocity. We reference energies to the global minimum $U_{min}$ and display points satisfying:

$$U(x, y) - U_{min} \leq \Delta U_{cut}$$

where $\Delta U_{cut}$ is chosen as a small multiple of $k_B T$ (e.g. $2\sim3\ k_B T$), retained for visualization and sampling. Equivalently we apply a probability cutoff $P_{cut}$ corresponding to the same energy window:

$$P(x, y) \propto \exp\left[-\, {U_{BL} + U_{pin} + U_L}\Big/{k_B T}\right] \geq P_{cut}$$

Using the characteristic vortex-line energy $\varepsilon = \frac{\Phi_0^2}{4\pi\mu_0\lambda^2} ln\kappa$, the bias-induced reduction integrated over the sampling layer is ~ $10^{-19}$ J per vortex[5–7], sufficient to bring edge regions into view. In the maps, red dots mark zero-bias entry/occupation sites; under bias, yellow dots indicate the expected drift (representative, no time integration). The four panel pairs (no-bias vs biased; asymmetric vs symmetric) thus show: (i) higher $U_{BL}$ and suppressed $P$ on the asymmetric edge at zero bias; and (ii) direction-dependent barrier lowering and earlier depinning on the symmetric edge under reverse bias, consistent with a diode-like response.

**Supplementary Note S5. Stray-field crosstalk analysis**

To assess possible interference between neighboring devices, we performed MuMax3 micromagnetic simulations of the stray-field distribution near the asymmetric edge of the Co layer. The simulated out-of-plane stray field $B_z$ decays rapidly with distance from the device edge, falling below ~1 Oe at 4 µm and below 0.55 Oe beyond 6 µm. This short decay length indicates that, for typical device spacings of several micrometers, stray-field-induced crosstalk is negligible.

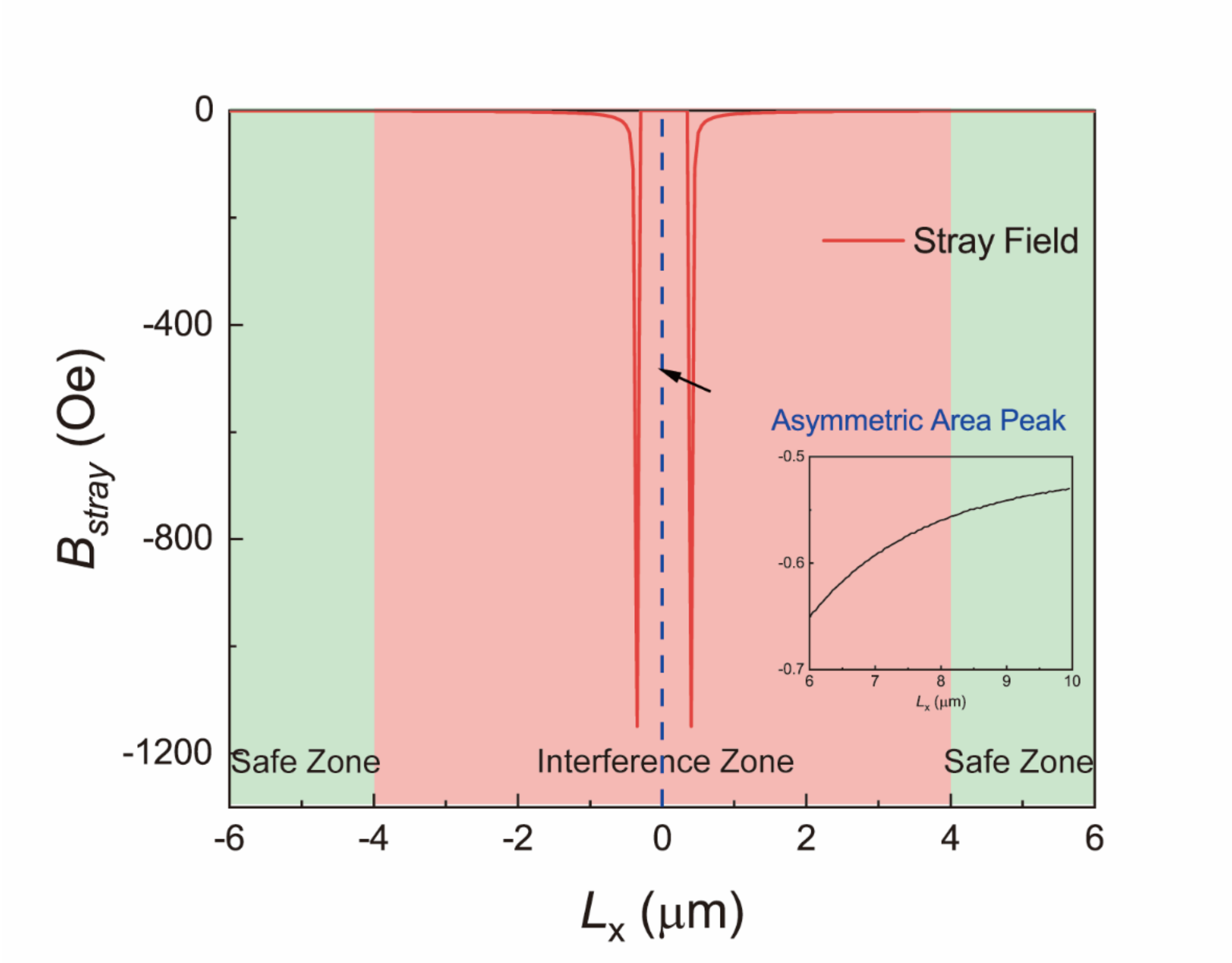

Figure S5. Simulated spatial decay of the out-of-plane stray field $B_z$ near the asymmetric Co edge. The field magnitude decreases rapidly and becomes negligible beyond a few micrometers from the device boundary.